\documentclass[12pt]{iopart}

\usepackage{color}

\usepackage{graphicx}

\begin{document}

\title{Passive scalar cascades in rotating helical and non-helical flows}

\author{P Rodriguez Imazio$^{1}$ and P D Mininni$^{1,2}$}
\address{$^1$ Departamento de F\'\i sica, Facultad de Ciencias Exactas y
         Naturales, Universidad de Buenos Aires and CONICET, Ciudad Universitaria, 1428
         Buenos Aires, Argentina. \\
             $^2$ NCAR, P.O. Box 3000, Boulder, Colorado 80307-3000, U.S.A.}
\ead{paolaimazio@df.uba.ar}

\begin{abstract}
We study how helicity affects the spectrum of a passive scalar in 
rotating turbulent flows, using numerical simulations of turbulent 
flows with or without rotation, and with or without injection of 
helicity. Scaling laws for energy and passive scalar spectra in the 
direction perpendicular to the rotation axis differ in rotating
helical flows from the ones found in the non-helical case, with 
the spectrum of passive scalar variance in the former case being 
shallower than in the latter. A simple phenomenological model that 
links the effects of helicity on the energy spectrum with the 
passive scalar spectrum is presented. 
\end{abstract}
\maketitle


\section{Introduction}

Enhanced mixing and transport are some of the most important 
properties of turbulent flows. These properties, sometimes
characterized by a turbulent diffusivity, result in rapid 
homogenization of any mixture of different fluids, are used for
many applications \cite{Falkovich, Rizzo}, 
and are also relevant in 
many atmospheric and oceanic flows \cite{Takemi}. 
In many of these flows rotation is important, and it is widely accepted that turbulent 
mixing is affected by the presence of rotation 
\cite{Vassilicos, Cambon 2004, Brandenburg}.

Several studies consider the effect of rotation in the energy
cascade. While the energy still undergoes a direct cascade, there is
evidence that at moderate rotation rates a fraction of the energy can
also undergo an inverse energy cascade, resulting in accumulation of
energy at scales larger than the energy injection scale
\cite{Waleffe, Smith}. 
Nowadays, it is also known that the presence of rotation
sets a preferential direction for the transfer of energy in spectral
space, with the energy going towards modes with small parallel
wavenumber (where parallel is defined relative to the rotation axis),
and resulting in a quasi-bidimensionalization of the flow
\cite{Waleffe, Cambon 1989, Cambon 1997}. The energy 
flux is also reduced (when compared with the homogeneous and 
isotropic case) per virtue of the extra resonance (or quasi-resonance) 
condition that triads must fulfil for the coupling between modes to
be effective \cite{Waleffe, Cambon 1997}. This results in a
steeper energy spectrum than the one expected from Kolmogorov
phenomenology. The effect of helicity in rotating turbulence has 
received less attention, although it is known that helicity is
relevant in many atmospheric processes, such as convective thunderstorms 
\cite{lilly, Kerr, Markowsky}, and it is also known to be important in
flows in blood vessels \cite{Rizzo}. For this latter case, results in 
\cite{mininni 2009} indicate that helicity affects the energy transfer 
to smaller scales, making the energy spectrum even steeper than in the
rotating non-helical case.

A paradigmatical way to study turbulent mixing is to consider the
advection and diffusion of a passive scalar by a turbulent velocity
field. When the flow is turbulent, mixing and transport of a scalar
quantity (such as density of pollutants or aerosols) is greatly
enhanced. The turbulent diffusion of a passive scalar in two-point
closures is related to the amplitude of the velocity
turbulent fluctuations \cite{Chkhetiani}, and therefore 
it can be expected that changes in the scaling law followed by the 
energy spectrum should affect the dynamics of the passive scalar. 
The scaling (including intermittency) of passive scalars in isotropic 
and homogeneous turbulent flows was studied in 
\cite{Kraichnan 1968}, and later in 
\cite{Falkovich,Sreenivasan,kraichnanychen,warhaft}. The 
Kraichnan model \cite{Kraichnan 1994} allowed computation of 
all scaling exponents of the passive scalar for a random, 
delta-correlated in time velocity field. The predictions are in good 
agreement in results from numerical simulations \cite{kraichnanychen}, 
which obtained a joint cascade of energy and passive scalar variance 
following the same scaling law given by the Kolmogorov spectrum 
\cite{Kolmogorov}, except for intermittency corrections.
 
Passive scalars in rotating turbulence have also been studied in 
numerical simulations, showing that the transport is affected by 
rotation and anisotropy \cite{yeung,brethouwer}. Recent numerical 
studies or rotating turbulence \cite{imazio 2011} show that
passive scalar variance is transferred preferentially toward modes 
with small parallel wavenumbers (i.e., quasi-bidimensional modes), 
following an inertial range scaling consistent with the 
bidimentionalization of the flow. Furthermore, the results show
that perpendicular structure functions of the passive scalar have 
anomalous scaling consistent with the Kraichnan model in a two 
dimensional (2D) space, again indicating strong anisotropic mixing 
and transport of scalar quantities in rotating flows. Experimental 
evidence of anomalous scaling of passive scalar structure functions 
in rotating flows was also observed in \cite{moisy}. 

Stochastic models and two-point closures indicate that two-particle 
dispersion in rotating turbulent flows is highly anisotropic, with 
different dispersion in the direction parallel and perpendicular to
the rotation axis \cite{Vassilicos,Cambon 2004,Kimura}, which can be related to 
the diffusion of passive scalars. Numerical simulations 
\cite{Brandenburg} also found that the turbulent diffusivity 
becomes anisotropic with rotation, reducing horizontal transport to 
a much lesser extend than vertical transport.

As for the case of the effect of helicity in the transport and mixing
of passive scalars, it was showed in \cite{Moffat 1983} that helicity 
affects passive scalar diffusivity in a turbulent flow. For isotropic
and homogeneous turbulence, it is argued that the lack of reflectional 
symmetry (related with a non zero value of the helicity) produces a 
turbulent skew-diffusion perpendicular to the local mean scalar 
gradient. Later, it was shown in \cite{Chkhetiani} using
renormalization groups that while anomalous scaling 
of the passive scalar is not affected by helicity, the turbulent
diffusion is. However, the effect of helicity in the transport of
scalar quantities in rotating helical flows has not been considered.

The aim of this paper is to study how helicity affects the spectrum
of a passive scalar in a rotating turbulent flow. The spectrum and flux 
are studied in numerical simulations of turbulent flows with or 
without rotation, and with or without injection of helicity, to 
identify spectral indices in the inertial range of the direct energy
and passive scalar cascades. The simulations are done with a 
parallel pseudospectral code with periodic boundary conditions 
\cite{Gomez05a,Gomez05b}, using a spatial resolution of $512^{3}$ 
grid points. Forcing used for all fields is a superposition of random 
modes, delta-correlated in time, with controllable helicity
injection, which in the simulations presented here is either zero or maximal.

Scaling laws for energy and passive scalar spectra in the direction 
perpendicular to the rotation axis differ in rotating helical flows 
from the ones found in the non-helical case. A phenomenological 
argument that links the effects of helicity on the energy spectrum 
with the passive scalar spectrum is also presented. 

\section{Equations and numerical simulations}

The data analized in the following section is obtained from direct 
numerical simulations of the incompressible Navier-Stokes equations 
for the velocity field ${\bf u}$ together with the equation for the 
passive scalar $\theta$, given by
\begin{equation}
\partial_t {\bf u} + {\bf u}\cdot \nabla {\bf u} = -2{\bf \Omega} \times 
    {\bf u} - \nabla p + \nu \nabla^2 {\bf u} +{\bf f},
\label{eq:NS}
\end{equation}
\begin{equation}
\nabla \cdot {\bf u} =0, 
\label{eq:incompressible}
\end{equation}
\begin{equation}
\partial_t {\theta} + {\bf u}\cdot \nabla {\theta} =  
    \kappa \nabla^2 {\theta} +{\bf \phi},
\label{eq:theta}
\end{equation}
where $p$ is the pressure divided by the (uniform) mass density, 
$\nu$ is the kinematic viscosity, and $\kappa$ is the scalar
diffusivity. Here, ${\bf f}$ is an external force that drives the 
turbulence, ${\phi}$ is the source of the scalar field, and 
${\bf \Omega} = \Omega \hat{z}$ is the rotation.

The numerical code used to solve Eqs.~(\ref{eq:NS})-(\ref{eq:theta}) 
in a three dimensional domain of size $2\pi$ with periodic boundary 
conditions is a second-order in time pseudospectral code, 
parallelized using the Message Passing Interface (MPI) library and 
OpenMP \cite{Gomez05a,Gomez05b, mininni 2011}. 
To solve the pressure, we take the divergence 
of Eq.~(\ref{eq:NS}), use the incompressibility condition 
(\ref{eq:incompressible}), and solve the resulting Poisson equation. 
To evolve in time a Runge-Kutta method with low storage is used. The 
code uses the $2/3$-rule for dealiasing, and as a result the maximum 
resolved wave number is $k_{max} = N/3$, where $N=512$ is the 
linear resolution. All simulations presented are well resolved, in the 
sense that the dissipation wave numbers $k_\nu$ and $k_\kappa$ 
(respectively for the kinetic energy and for the passive scalar) are
smaller than the maximum wave number $k_{max}$ at all times.

Dimensionless parameters used to control the simulations are the 
Reynolds $R_e$, the Peclet $P_e$, and the Rossby  $R_o$ numbers, 
respectively given by
\begin{equation}
R_e=\frac{UL}{\nu},
\end{equation}
\begin{equation}
P_e=\frac{UL}{\kappa},
\end{equation}
and
\begin{equation}
R_o = \frac{U}{2L\Omega},
\end{equation}
where $U$ is the root mean square velocity, and $L$ is the flow 
forcing scale defined as $L=2\pi/k_F$ with $k_F$ the forcing 
wave number. For most of the simulations shown in the 
following section $U \approx 2$, and all runs have 
$\nu=\kappa$ (i.e., $P_e=R_e$). The forcing used for the 
velocity field as well as for the passive scalar is a 
superposition of Fourier modes with random phases, 
delta-correlated in time, and the amount of helicity injected is 
controlled by correlating the velocity field components and the 
phases between Fourier modes using the method described in 
\cite{Patterson}.

Both the kinetic energy and the passive scalar variance were 
injected at the same wave number $k_F$. One set of runs (set A) 
has external forcing applied at $k \in [1,2]$ (therefore 
$k_F\approx 1$, and the simulations have the largest possible 
separation of scales between the forcing wavenumber and the 
largest resolved wavenumber). Another set of runs (set B) has 
forcing at $k \in [2,3]$ (then $k_F \approx 2$). Finally, a third 
set of runs is forced at $k_F=3$ (set C). For the last set of runs, 
the choice of $k_F=3$ results in a small separation of scales 
between the box size and the forcing scale, allowing for some 
of the energy to be transferred to larger scales in the presence 
of rotation (although the separation of scales is not large enough 
to study the inverse cascade). This reduces the Reynolds number 
and results in a narrower direct cascade inertial range, since the 
separation between the forcing and the dissipation scale is 
reduced. However, the incipient inverse transfer of energy that 
develops is important for the development of a dominant direct 
cascade of helicity when helicity is injected in the presence of 
rotation (see \cite{mininni 2009}). The three 
sets of runs allow us to compare runs with similar Rossby numbers 
(albeit with different Reynolds numbers) while varying the 
amount of helicity. While the flows in set B are helical, the flows 
in sets A and C are non-helical (more details of the runs in sets 
A and C can be found in \cite{imazio 2011}).

The simulations were performed as follows: first a simulation of the 
Navier-Stokes equation with $\Omega=0$ was done, until a turbulent 
steady state was reached (this requires an integration for
approximately ten turnover times). Then the passive scalar was
injected, and the run was continued for other ten turnover times 
until a steady state for the passive scalar was reached (these runs 
correspond to runs A1, B1, and C1). Finally, rotation was 
turned on. Different values of $\Omega$ were considered to have 
similar Rossby numbers in all the runs with $\Omega \neq 0$. All 
the runs with rotation in each set were started using as initial 
conditions for the velocity and the passive scalar the latest output of 
the runs without rotation in the same set (runs A1, B1, or C1 respectivelly). Each 
of the runs with rotation was continued for over twenty turnover 
times. Parameters for all runs are listed in Table~$\bf 1$.


\begin{center}
\tabcolsep=3pt  
\small
\renewcommand\arraystretch{1.2}  
\begin{minipage}{15.5cm}{
\small{\bf Table 1.} Parameters used in the simulations: $k_F$ is the forcing 
wavenumber, $\Omega$ is the rotation angular velocity, $R_o$ is the
Rossby number, $\nu$ the kinematic viscosity, $R_e$ the Reynolds
number, $U$ the r.m.s. velocity in the turbulent steady state, and 
$H=\left<{\bf u}\cdot \nabla \times {\bf u}\right>$ is the total
helicity (averaged in time).}
\end{minipage}
\vglue5pt

\begin{tabular}{| c | c | c | c | c | c | c | c | }  
\hline 
{Run} & {$k_{F}$}&{$\Omega$} & {$R_{o}$} & {$\nu$} & {$R_e$} & {$U$} & {$H$} \\   
\hline
{A1} & {1} & {$0$} & {$\infty$} & {$6\times 10^{-4}$} & {$1000$} & {$2$} & {$0$} \\   
{A2} & {1} & {$4$} & {$0.04$} & {$6\times 10^{-4}$} & {$1000$} & {$2$} & {$0$} \\   
{B1} & {2} & {$0$} & {$\infty$} & {$5\times 10^{-4}$} & {$600$} & {$2$} & {$3$} \\   
{B2} & {2} & {$8$} & {$0.04$} & {$5\times 10^{-4}$} & {$600$} & {$2$} & {$6$} \\   
{B3} & {2} & {$16$} & {$0.04$} & {$5\times 10^{-4}$} & {$600$} & {$4$} & {$11$} \\   
{C1} & {3} & {$0$} & {$\infty$} & {$6\times 10^{-4}$} & {$240$} & {$2$} & {$0$} \\   
{C2} & {3} & {$12$} & {$0.04$} & {$6\times 10^{-4}$} & {$240$} & {$2$} & {$0$}\\   
\hline
\end{tabular}
\end{center}
\vspace*{2mm}

\section{Numerical results}

\subsection{Effect of rotation}

\begin {figure}
\begin{center}
\includegraphics[width=8.3cm]{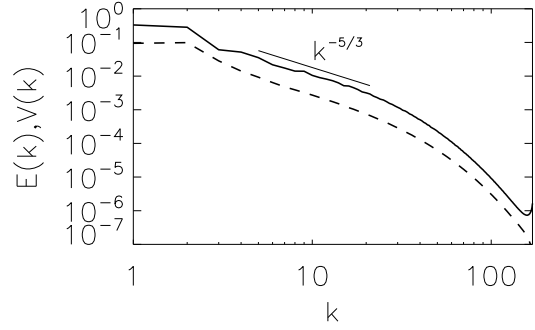}
\end{center}
\caption {Isotropic energy (solid line) and passive scalar 
(dashed line) spectrum for run A1 (without rotation and 
without helicity injection). Kolmogorov scaling is indicated 
as a reference.}
\label{fig:fig1}
\end{figure}

Figure \ref{fig:fig1} shows the isotropic energy $E(k)$ and passive 
scalar variance $V(k)$ spectra for run $A1$ (without rotation and 
without helicity injection). An inertial range can be identified,
where energy and passive scalar follow a $k^{-5/3}$ scaling law, as 
expected from previous studies of passive scalar in isotropic and 
homogeneous turbulence \cite{Falkovich,Sreenivasan, kraichnanychen}. Runs B1 and C1, 
also without rotation but forced at different wavenumbers (and 
in the case of run B1, with helicity) show the same scaling. 

\begin {figure}
\begin{center}
\includegraphics[width=8.3cm]{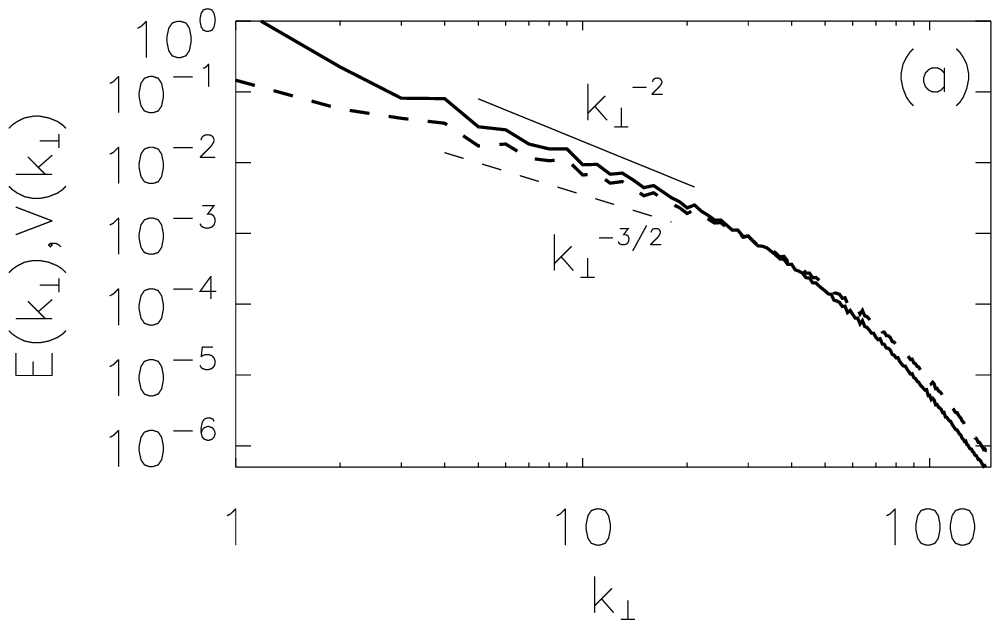}\\
\includegraphics[width=8.3cm]{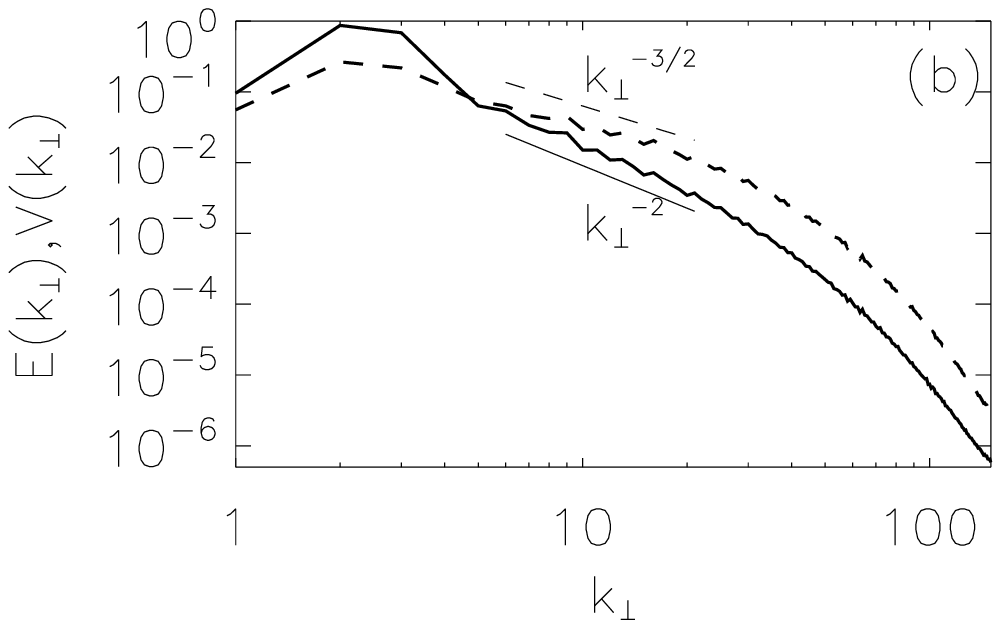}
\end{center}
\caption {(a) Reduced perpendicular spectrum for the energy (solid 
line) and for the passive scalar variance (dashed line) for (a) run
A2 ($\Omega=4$), and  (b) run C2 ($\Omega=12$). 
Scaling laws $\sim k_\perp^{-2}$ and $\sim k_\perp^{-3/2}$ are 
indicated as references.}
\label{fig:fig2}.
\end{figure}

In Fig.~\ref{fig:fig2} we show the energy and
passive scalar reduced perpendicular spectra, respectively
$E(k_\perp)$ and $V(k_\perp)$, for runs $A2$ and $C2$ 
(corresponding to flows with rotation but without net helicity, 
see table $1$). The reduced perpendicular 
spectrum is obtained by summing over all wavenumbers in 
Fourier space in cylindrical shells with radius 
$k_{\perp} = \sqrt{(k_{x}^2+k_{y}^2)}$, to take into account the 
fact that the flows become anisotropic in the presence of 
rotation (see \cite{Cambon 1997,mininni 2012} for 
definitions and details of anisotropic spectra).

\begin{figure}
\begin{center}
\includegraphics[width=8.3cm]{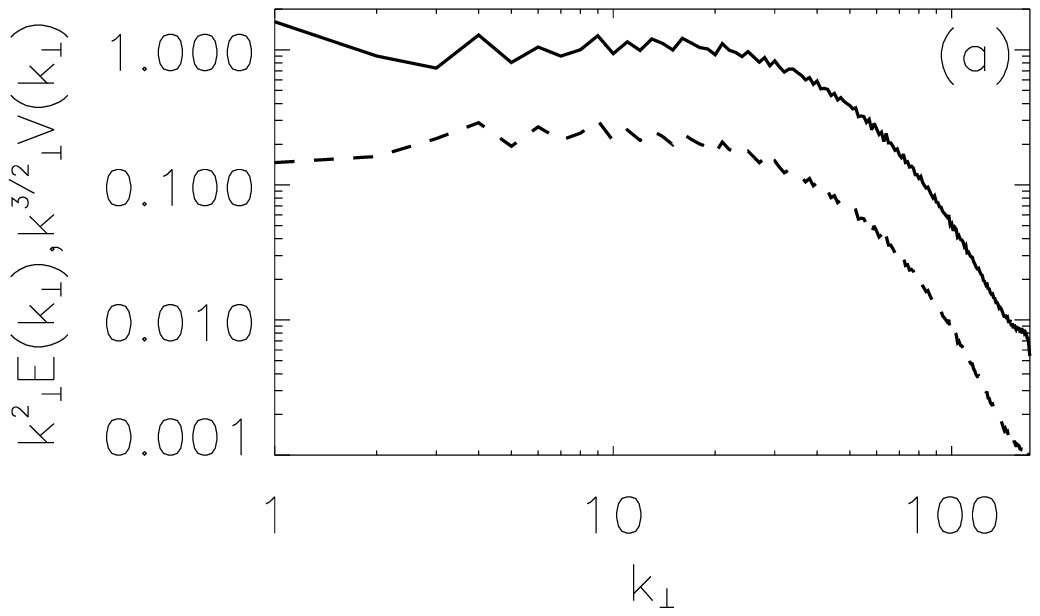}\\
\includegraphics[width=8.3cm]{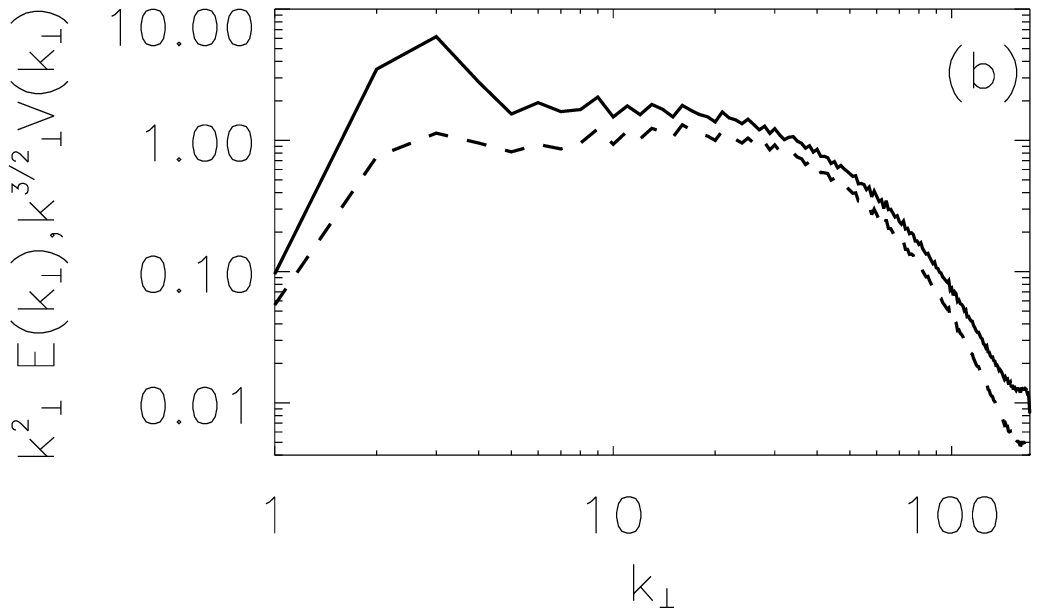}
\end{center}
\caption {(a) Reduced perpendicular spectrum for the energy (solid 
line) and for the passive scalar variance (dashed line) compensated
respectivelly by  $-2$ and $-3/2$ in run
A2. (b) The same for run C2.}
\label{fig:fig3}.
\end{figure}

As can be seen in Figs.~\ref{fig:fig2}(a) and \ref{fig:fig2}(b), inertial 
range scaling can be identified for both the energy and the passive 
scalar variance, although with different power laws. The reduced 
perpendicular energy spectrum follows a $\sim k_{\perp}^{-2}$ 
scaling. This power law has been already reported in numerical 
simulations and experiments of rotating turbulence (see, e.g., 
\cite{Muller, mininni 2009}), and is consistent with simple 
phenomenological models based on a slow down of the energy 
transfer associated with the interaction between waves and eddies 
\cite{Zhou, Muller, mininni 2009}, as well as with more 
detailed two-point closures \cite{Cambon 1989,Cambon 1997}. 
The passive scalar inertial range displays a scaling compatible 
with $\sim k_{\perp}^{-3/2}$ scaling, as also reported in 
\cite{imazio 2011}. These power laws can be further confirmed 
when the spectra are compensated (see Figs.~\ref{fig:fig3}(a) and 
\ref{fig:fig3}(b)). Unlike the case of isotropic and homogeneous 
turbulence, in the presence of rotation the kinetic energy and 
the passive scalar show different power laws in the inertial range. 

\begin{figure}
\begin{center}
\includegraphics[width=8.3cm]{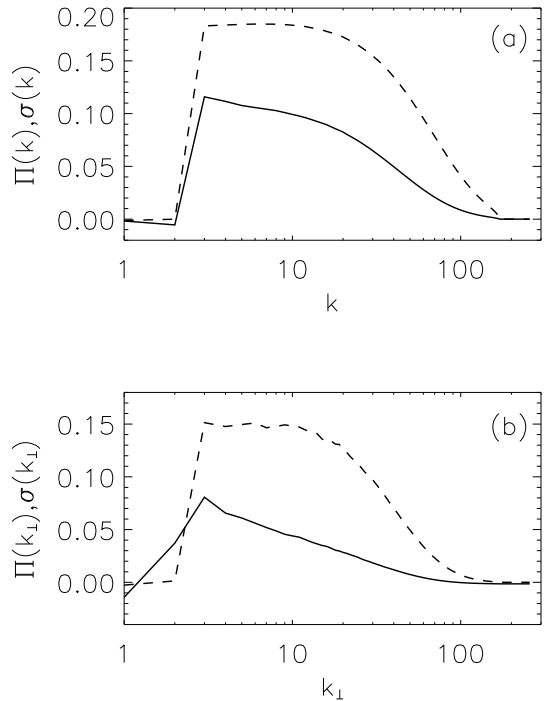}\\
\end{center}
\caption{(a) Energy flux $\Pi(k)$ (solid line) and passive scalar flux 
$\sigma(k)$ (dashed line) for the run without rotation C1. (b) 
Perpendicular energy flux $\Pi(k_\perp)$
and perpendicular passive scalar flux $\sigma(k_\perp)$ for the run 
with rotation C2.}
\label{fig:fig4}
\end{figure}

The inertial ranges indicated in Figs.~\ref{fig:fig2} and \ref{fig:fig3}
correspond to direct cascades of energy and scalar variance. This can 
be confirmed from the energy and passive scalar spectral fluxes shown
in Fig.~\ref{fig:fig4} for runs C1 and C2 (respectively without and 
with rotation). In the non-rotating case, energy shows a range
of approximately constant (and positive) flux, indicating energy is 
transferred toward smaller scales, while the energy flux is negligible 
for wave numbers smaller than the forcing wave number 
($k < k_{F} = 3$). The passive scalar variance also direct cascades to 
smaller scales with a range of wave numbers with approximately 
constant flux. When rotation is present, the energy flux becomes 
negative for $k < k_{F}$ (indicating a fraction of the energy is 
transferred towards scales larger than the forcing scale, although
without enough scale separation to develop an inverse cascade), 
while the energy flux towards smaller scales remains positive 
although it decreases when compared with run C1. For the passive 
scalar, no significative flux toward larger scales is observed, and 
the cascade remains direct with also a small decrease of the 
positive (direct) flux for $k>k_F$ when compared with non-rotating
case.

\begin{figure}
\begin{center}
\includegraphics[width=9.3cm]{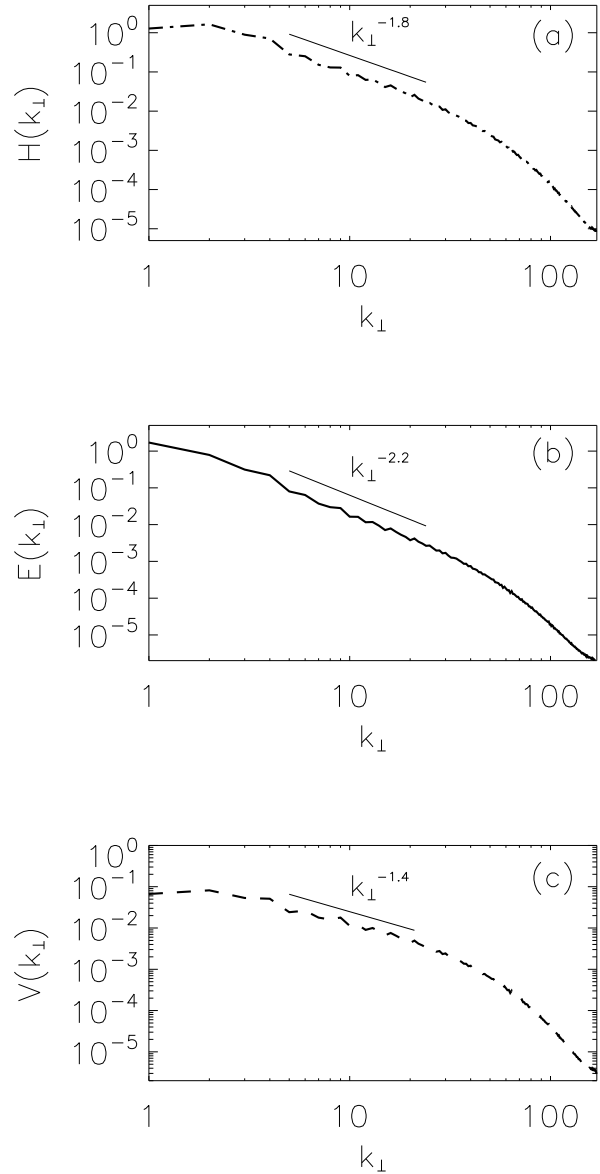}
\end{center}
\caption{(a) Reduced perpendicular helicity spectrum, (b) energy 
spectrum, and (c) passive scalar spectrum for run B2 (helical
turbulent flow with $\Omega = 8$). In all cases slopes are 
indicated as references.}
\label{fig:fig5}
\end{figure}

\subsection{Effect of helicity}

\begin{figure}
\begin{center}
\includegraphics[width=9.3cm]{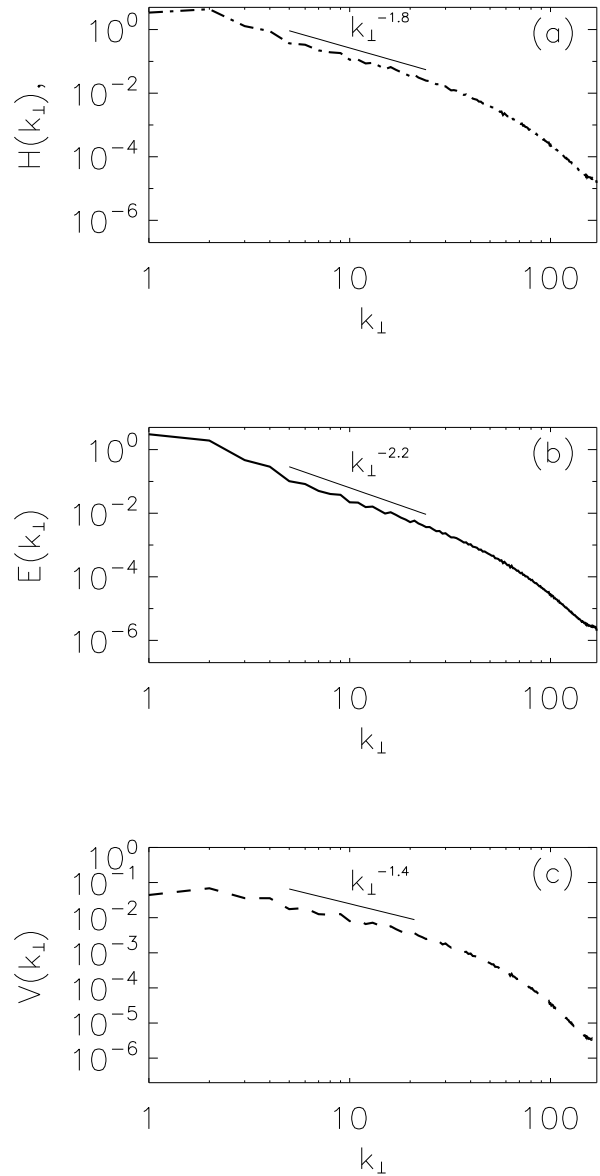}
\end{center}
\caption{(a) Reduced perpendicular helicity spectrum, (b) energy 
spectrum, and (c) passive scalar spectrum for run B3 (helical
turbulent flow with $\Omega = 16$). In all cases slopes are 
indicated as references.}
\label{fig:fig6}
\end{figure}

Now we analyze the runs with rotation and with maximal helicity
injection, resulting in anisotropic helical turbulent flows. Figures 
\ref{fig:fig5} and \ref{fig:fig6} show the helicity, energy, and 
passive scalar reduced perpendicular spectra for runs B2 and B3. 
Slopes with reference values 
for the scaling in the inertial range are also indicated. While without
rotation helicity does not change the scaling of the passive scalar 
spectrum, in the rotating case a difference is observed. A 
careful analysis of the spectrum indicates that the passive 
scalar is close to a $\sim k_{\perp}^{-1.4}$ power law, a spectrum 
slightly shallower than the one observed in runs A2 and C2. 
The shallower spectrum observed for $V(k_\perp)$ is associated with a 
change in the energy spectrum when helicity is present.

The energy spectrum in (helical) runs B2 and B3 is steeper than 
in the (non-helical) runs A2 and C2, as can be also seen in 
Figs.~\ref{fig:fig5} and \ref{fig:fig6}. The inertial ranges are 
compatible with a $\sim k_{\perp}^{-2.2}$ power law. This result 
is compatible with the results reported in \cite{mininni 2009}, 
where numerical simulations were presented showing that in 
rotating helical flows the direct flux of helicity dominates over 
the direct flux of energy, affecting the scaling law for the energy 
in the direct cascade range. A phenomenological argument was 
also presented, which assuming the direct cascade of helicity is 
dominant, results in a spectrum 
$E(k_\perp) H(k_\perp) \sim k_\perp^{-4}$. In other words, if the 
energy spectrum satisfies $E(k)\sim k^{-n}$, then the helicity 
should scale as $H(k)\sim k^{4-n}$; $n$ becomes larger (and the 
energy spectrum steeper) as the flow becomes more helical, 
with the limit $n = 2.5$ for the case of a maximally helical 
turbulent flow (in practice, this limit cannot be obtained, as a 
flow with maximal helicity has the non-linear term in the 
Navier-Stokes equation equal to zero, and therefore no transfer 
can take place).

\begin{figure}
\begin{center}
\includegraphics[width=8.3cm]{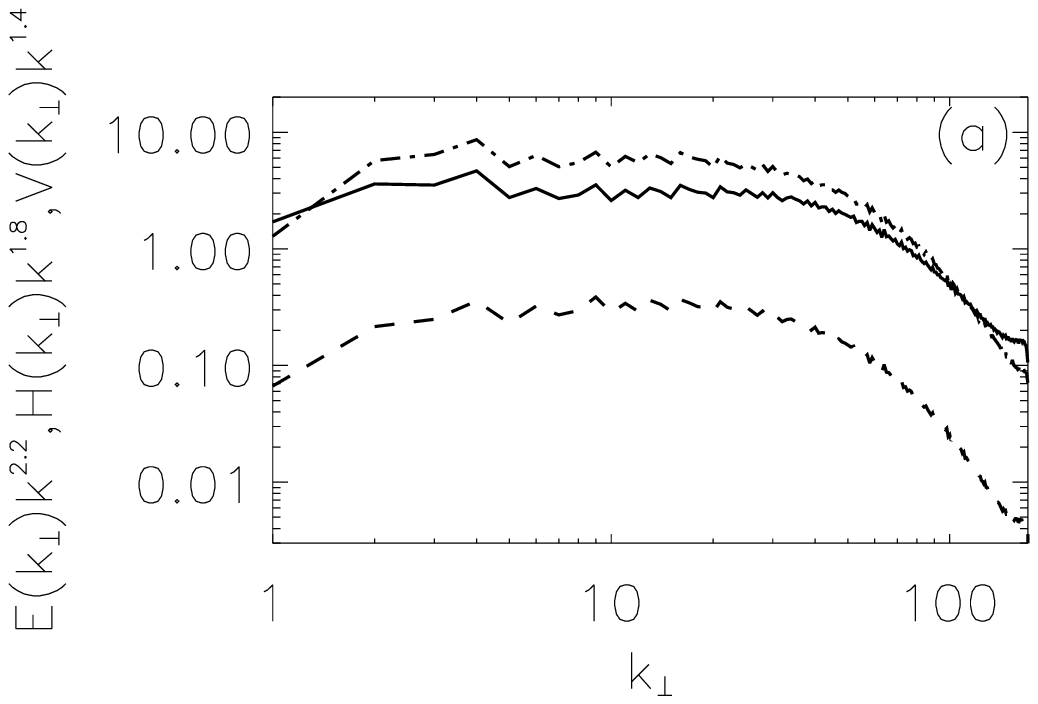}\\
\includegraphics[width=8.3cm]{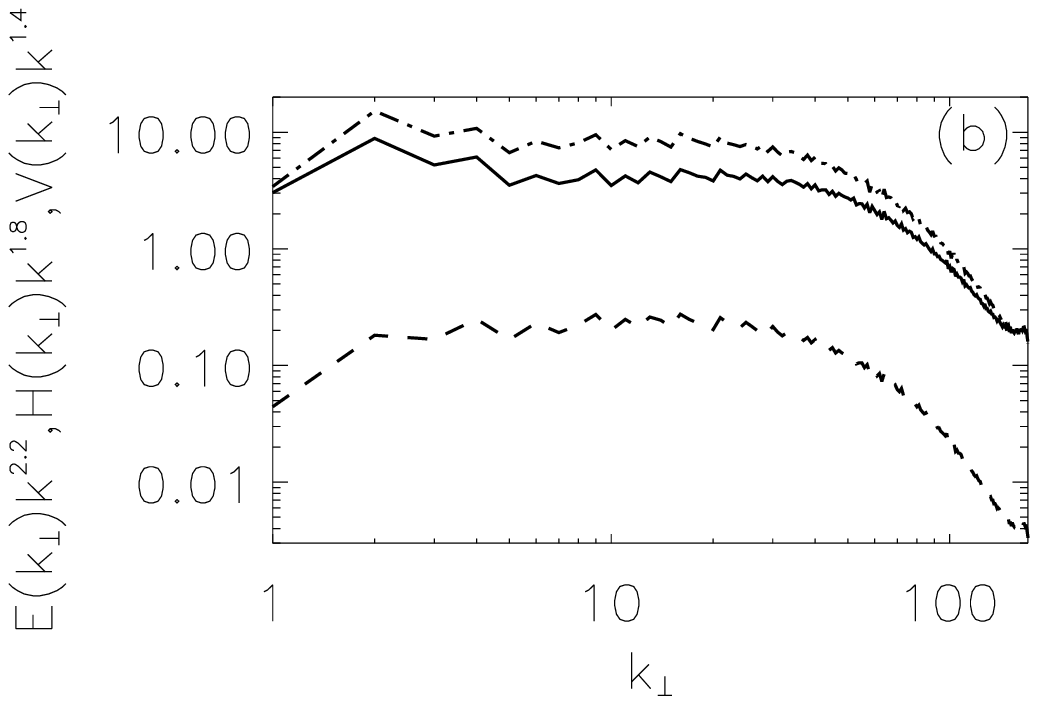}
\end{center}
\caption{Reduced perpendicular spectra for the helicity 
(dash-dotted line), energy (solid line), and passive scalar (dashed 
line) compensated respectively by $k_\perp^{-1.8}$, 
$k_\perp^{-2.2}$, and $k_\perp^{-1.4}$, in helical runs (a) B2 and 
(b) B3.}
\label{fig:fig7}
\end{figure}

The behavior of the helicity spectrum in runs B2 and B3 is 
consistent with the phenomenological argument described 
above. In Figs.~\ref{fig:fig5} and \ref{fig:fig6}, a scaling 
$\sim k_{\perp}^{-1.8}$ is indicated as a reference, which 
seems compatible with the behavior of $H(k_\perp)$. 
Compensated spectra for the energy, the helicity, and the 
passive scalar for runs B2 and B3 are shown in 
Fig.~\ref{fig:fig7}. A good agreement between the reference 
slopes and the numerical data is apparent.

Following the phenomenological argument mentioned above 
for the energy spectrum, we can put forward a simple argument to 
explain the difference observed in the scaling of the passive scalar 
in rotating helical and non-helical turbulent flows. From 
Eq.~(\ref{eq:theta}), it can be seen on dimensional grounds that
for scales in the inertial range, the passive scalar flux across the 
scale $l_\perp$ (equal to the passive scalar injection rate) 
$\sigma = \partial_{t} \left\langle \theta^{2} \right\rangle$ 
must be
\begin{equation}
\sigma \sim \frac{\theta_{l_\perp}^{2} u_{l_\perp}}{l_\perp},
\label{eq:sigma}
\end{equation}
where $\theta_{l_\perp}$ is the characteristic concentration of the 
passive scalar at the scale $l_\perp$, and $u_{l_\perp}$ the 
characteristic velocity (since the flow becomes anisotropic in the 
presence of rotation, we are assuming most of the fluctuations are 
concentrated in structures with weak variation in the direction 
along the axis of rotation). If $\sigma$ is constant in the inertial 
range, we can estimate the passive scalar spectrum 
$V(k_\perp) \sim \theta_{l_\perp}^{2}/k_\perp$ from
Eq.~(\ref{eq:sigma}) as
\begin{equation}
V(k_\perp) \sim \frac{\sigma l_\perp^{2}}{u_{l_\perp}}.
\label{eq:V1}
\end{equation}
If the energy spectrum is $E(k_\perp)\sim k_\perp^{-n}$, and 
therefore the characteristic velocity at a scale $l_\perp$ is 
$u_{l_\perp} \sim l_\perp^{1-n}$, the passive scalar spectrum results
\begin{equation}
V(k_\perp) \sim \sigma l_\perp^{\frac{5-n}{2}}\sim \sigma k_\perp^{- \frac{5-n}{2}}.
\label{eq:V1}
\end{equation}
Therefore, the spectral index for the passive scalar is given by 
$n_{\theta}=(5-n)/2$. This result is also valid in the isotropic case, 
provided $l_\perp$ is replaced by $l$.

\begin{figure}
\begin{center}
\includegraphics[width=8.3cm]{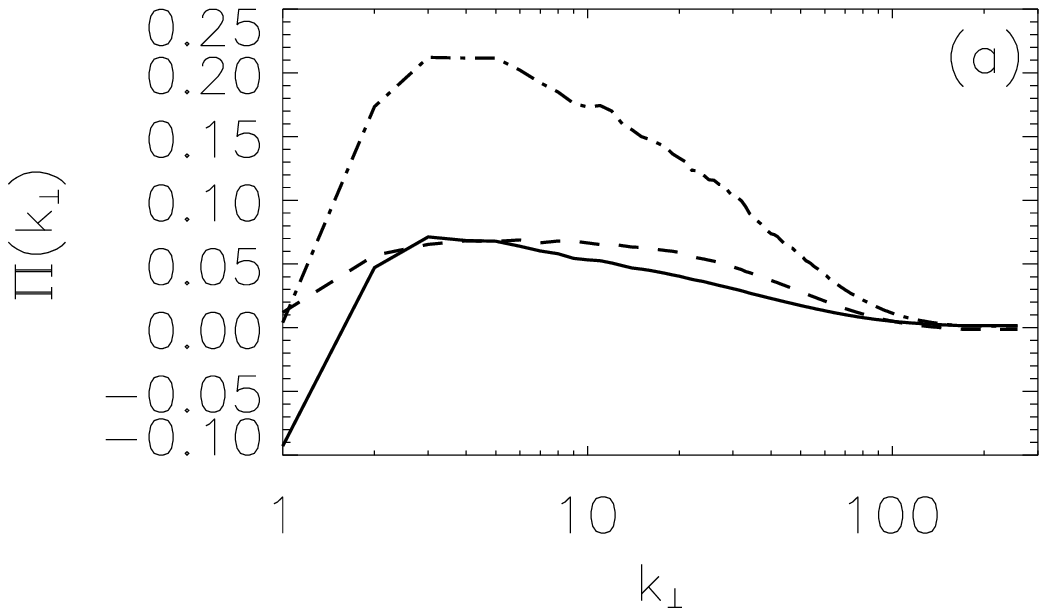}\\
\includegraphics[width=8.3cm]{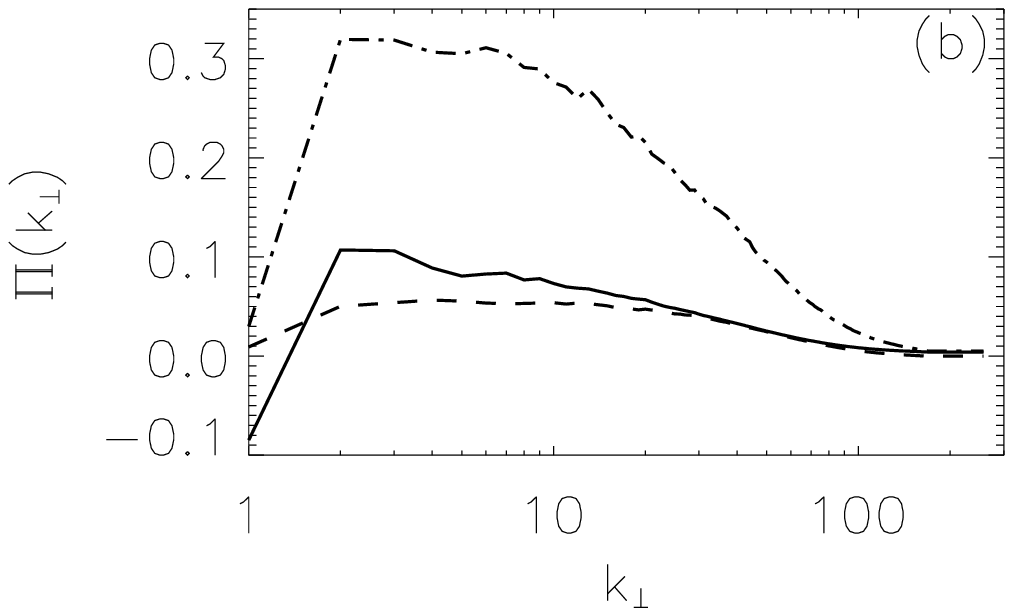}
\end{center}
\caption{Perpendicular helicity flux $\Sigma(k_\perp)/k_F$ (dash-dotted 
line), energy flux $\Pi(k_\perp)$ (solid line), and passive scalar flux 
$\sigma(k)$ (dashed) for runs (a) B2 and (b) B3. }
\label{fig:fig8}
\end{figure}

The numerical results are in good agreement with this simple 
phenomenological argument. If $n\approx 2$ (runs with rotation but 
without helicity), then $n_{\theta} \approx 3/2$. On the other hand, if 
$n\approx 2.2$ (compatible with the spectrum observed in the runs 
with rotation and helicity), then $n_{\theta} \approx 1.4$.

That the fluxes are still positive (i.e., the cascades direct) and 
approximately constant (within the limitations imposed by the 
spatial resolution and the moderate Reynolds numbers considered) 
in rotating helical flows can be confirmed from the helicity, energy, 
and passive scalar fluxes shown in Fig.~\ref{fig:fig8} for runs B2 
and B3 (the helicity flux in the figure is divided by $k_F$ to 
compare all fluxes with the same units). 
As in the runs without helicity, the energy 
flux shows some inverse transfer towards larger scales for $k<k_F$, 
while all other fluxes are positive everywhere indicating quantities 
are not transferred towards larger scales. An excess of helicity flux 
(when compared with the energy flux) can be observed, in 
agreement with the arguments of dominance of the helicity cascade 
in \cite{mininni 2009}.

We finish the analysis of the runs by quantifying the degree of 
anisotropy in the velocity field and in the passive scalar
distribution. As already mentioned, the presence of rotation 
results in a preferred transfer of energy towards two-dimensional 
modes. This motivated our study of the energy and passive scalar 
spectral scaling using the reduced perpendicular spectrum instead 
of the usual isotropic spectrum. We now quantify how much 
energy and passive scalar variance is in two-dimensional modes 
in each of the runs. Several anisotropy measures can be used to 
this end \cite{Cambon 1989, Cambon 1997,bartello}. As an 
example, the ratio of energy in all modes with $k_{\parallel}=0$ to 
the total energy, i.e., $E(k_\parallel =0)/E$, can be used to 
characterize large scale anisotropy \cite{mininni 2009}. For a 
purely two-dimensional flow, this ratio is equal to one. For the
passive scalar, the equivalent quantity $V(k_\parallel =0)/V$ can 
also be used. Finally, in helical flows we can also compute 
$H(k_\parallel =0)/H$ to quantify large scale anisotropy of the 
helicity.

\begin{center}
\tabcolsep=3pt  
\small
\renewcommand\arraystretch{1.2} 
\begin{minipage}{15.5cm}{
\small{\bf Table 2.} Anisotropy in helical and non-helical runs 
    with rotation. $E(k_\parallel =0)/E$ is the ratio of energy in all 
    modes with $k_{\parallel}=0$ to the total energy, 
    $V(k_\parallel =0)/V$ is the ratio of scalar variance in modes 
    with $k_{\parallel}=0$ to the total scalar variance, and 
    $H(k_\parallel =0)/H$ is the ratio of helicity in
    $k_{\parallel}=0$ modes to the total helicity. The angles 
    $\alpha_u$, $\alpha_\theta$, and $\alpha_H$ are respectively 
    the Shebalin angles for the velocity, the passive scalar, and the 
    helicity.}
\end{minipage}
\vglue5pt
\begin{tabular}{| c | c | c | c | c | c | c|}  
\hline 
{Run} & {$E(k_{\parallel})/E(k)$}&{$V(k_{\parallel)}/V(k)$} & {$H(k_{\parallel)}/H(k)$} & {$tan^{2} \alpha_{u}$} & {$tan^{2} \alpha_{\theta}$} & {$tan^{2} \alpha_{H}$}  \\   
 \hline
{A2} & {$0.5$} & {$0.4$}  & {$-$} & {$13$} & {$20$} & {$-$}  \\   
{B2} & {$0.6$} & {$0.25$}& {$0.27$} & {$17$} & {$37$}  & {$14$}  \\   
{B3} & {$0.4$} & {$0.24$}& {$0.17$} & {$18$} & {$76$}& {$20$} \\   
{C2} & {$0.2$} & {$0.1$} & {$-$} & {$14$} & {$50$}  & {$-$} \\   
\hline
\end{tabular}
\end{center}

As can be seen in Table~$2$, in all runs a substantial 
fraction of the energy, the passive scalar variance, and (to a lesser 
extent) the helicity, is in two-dimensional modes. Helicity does not 
seem to affect the large-scale anisotropy. Independently of the 
helicity in the flow the energy is more anisotropic at large scales 
than the passive scalar, as already found for non-helical rotating 
flows in \cite{imazio 2011}.

To characterize small scale anisotropy, the Shebalin angles can be 
used \cite{shebalin,matthaeus}. For the velocity field, the Shebalin 
angle is defined as
\begin{equation}
\tan^{2}(\alpha_u)= 2\lim_{l\to0}\frac{S_2(l_\perp)}{S_2(l_\parallel)}=
  2 \sum_{k_\perp}k_{\perp}^{2} E(k_\perp) \big/ \sum_{k_\parallel}k_{\parallel}^{2} E(k_\parallel),
\end{equation}
where $S_2(l_\parallel)$ and $S_2(l_\perp)$ are the second order
longitudinal structure functions of the velocity, respectively with 
spatial increments in the direction parallel and perpendicular to 
the axis of rotation. The angle $\alpha_{u}$ gives a global 
measure of small scale anisotropy, with a value of 
$\tan^{2}(\alpha_{u})=2$ corresponding to an isotropic flow, and 
larger values corresponding to more anisotropic flows. The 
definition is easily extended to the cases of the passive scalar 
and the helicity. Table~{2} shows the square 
tangent of the Shebalin angles for the velocity field 
($\tan^{2}(\alpha_u)$), for the passive scalar 
($\tan^{2}(\alpha_{\theta})$), and for the helicity 
($\tan^{2}(\alpha_H)$). At the small scales, the flows with helicity 
seem to develop stronger anisotropies for the passive scalar.

These quantities give information only on the global anisotropy 
of the velocity field and of the passive scalar. There are other 
ways to quantify spectral anisotropy, that give detailed 
information of the distribution of energy in spectral space, 
and of the degree of anisotropy at different scales, as the
axisymmetric spectrum $e(k_\perp,k_\parallel)$ \cite{Cambon 1997}. 
A detailed study of spectral anisotropy is left for 
future work.

\vspace*{2mm}

\section{Concluding remarks}

We presented preliminary results of numerical simulations of 
passive scalar advection and diffusion in rotating turbulent 
flows with and without helicity, in grids of $512^3$ points. 
 
While in isotropic and homogeneous turbulence at moderate 
Reynolds number the energy and the passive scalar variance 
follow Kolmogorov scaling $\sim k^{-5/3}$ except for
intermittency corrections, in the presence of 
rotation non-helical flows display a reduced perpendicular 
energy spectrum $E(k_\perp) \sim k_{\perp}^{-2}$ and a 
shallower reduced perpendicular spectrum 
$V(k_\perp) \sim k_{\perp}^{-3/2}$ for the passive scalar.

In the absence of rotation, the scaling of the energy and of 
the passive scalar remains the same independently of the 
level of helicity in the flow. In helical rotating flows, our 
simulations display a steeper energy spectrum compatible 
with $E(k_\perp) \sim k_{\perp}^{-2.2}$, and a shallower 
passive scalar spectrum compatible with 
$V(k_\perp) \sim k_{\perp}^{-1.4}$. These numerical results 
are consistent with a simple phenomenological model 
that predicts that if the energy spectrum has an inertial 
range of the form $E(k_\perp)\sim k_\perp^{-n}$, then the 
passive scalar spectrum follows a power law 
$V(k_\perp)\sim k_\perp^{-n_\theta}$ with spectral index 
$n_{\theta}=(5-n)/2$.

Finally, analysis of global measures of anisotropy indicate 
that the distribution of the passive scalar at small scales 
becomes more anisotropic in helical rotating flows (in 
comparison with the results in non-helical rotating flows)
but it is largely unaffected at large scales. 

The results open new questions that will be addressed in 
future works. In particular, and as the spectral scaling of the 
passive scalar in rotating flows seems to be affected by 
helicity, one may ask: Is intermittency and anomalous 
scaling of the passive scalar changed by helicity? And how 
is the transport and mixing of the passive scalar affected? 
While the former question can be answered by computing 
scaling exponents for rotating flows with and without 
helicity, the latter may require quantification of the 
turbulent transport in directions parallel and perpendicular 
to the axis of rotation.

\section*{Acknowledgments}
PDM acknowledges support from the Carrera del Investigador 
Cient\'{\i}fico of CONICET. The authors acknowledge support 
from grants UBACYT 20020090200692, PICT 2007-02211, 
and PIP 11220090100825.

\end{document}